\documentclass[11pt]{article}
\usepackage[utf8]{inputenc}
\usepackage{amsmath,amssymb,hyperref}
\usepackage{indentfirst}
\usepackage{graphicx}
\usepackage{multirow}
\usepackage{authblk}
\usepackage{float}
\usepackage{setspace}
\usepackage[margin=1.5in]{geometry}
\usepackage[authoryear]{natbib} 
\usepackage{xcolor}
\usepackage[normalem]{ulem}
\graphicspath{ {./fig/} }

\newcommand{\stkout}[1]{\ifmmode\text{\sout{\ensuremath{#1}}}\else\sout{#1}\fi}

\title{Forecasting Volatility with Machine Learning and Rough Volatility: Example from the Crypto-Winter}

\author{Siu Hin Tang\footnote{Department of Mathematics, National University of Singapore; email: s.h.tang@u.nus.edu}, Mathieu Rosenbaum\footnote{CMAP, Ecole Polytechnique email: mathieu.rosenbaum@polytechnique.edu}, and Chao Zhou\footnote{Department of Mathematics and Risk Management Institute, National University of Singapore; email: matzc@nus.edu.sg}}

\date{January 21, 2024}

\begin{document}

\maketitle

\textbf{Abstract.} We extend the application and test the performance of a recently introduced volatility prediction framework encompassing LSTM and rough volatility. Our asset class of interest is cryptocurrencies, at the beginning of the ``crypto-winter'' in 2022. We first show that to forecast volatility, a universal LSTM approach trained on a pool of assets outperforms traditional models. We then consider a parsimonious parametric model based on rough volatility and Zumbach effect. We obtain similar prediction performances with only five parameters whose values are non-asset-dependent. Our findings provide further evidence on the universality of the mechanisms underlying the volatility formation process.\\

\textbf{\textit{Keywords}} -- Machine learning, LSTM, rough volatility, quadratic rough Heston, Zumbach effect, cryptocurrencies, Bitcoin

\section{Introduction}

In \cite{rosenbaum_universality_2022}, classical parametric models, non-parametric Long Short-Term Memory (LSTM), as well as a novel parsimonious rough volatility-based predictor are used to predict stocks' volatility. The authors are successful in establishing some universal mechanism in the volatility formation process, and the rough volatility-based framework of \cite{rosenbaum_universality_2022} is able to forecast volatility better than traditional methods. In this paper, we are thus interested in testing the applicability and robustness of this approach by adapting the volatility forecasting framework to a different class of asset. Here we choose cryptocurrencies which is a much newer and younger market. Serving as a companion paper to \cite{rosenbaum_universality_2022}, we aim at answering two questions in this work:
\begin{itemize}
    \item Whether the devices used in the original paper predict volatility well for cryptocurrencies at the beginning of the so-called ``crypto-winter''.
    \item Whether the volatilities have the same properties as observed for stocks, such as the universality of their formation process across coins during this period.\\
\end{itemize}

Bitcoin, proposed in \cite{nakamoto_bitcoin_2008}, is a decentralized payment system where transactions are stored with blockchain technology, leveraging cryptographic algorithms. Many similar systems were introduced and formed a completely new asset class known as cryptocurrencies. Such electronic currencies offer many unique advantages over traditional fiat money. Since conception, cryptocurrencies have been a popular research topic in the academic community and have also received a lot of interest from financial institutions and the general public alike \citep{donier2015million,griffin2020bitcoin,arnosti2022bitcoin,bianchi_performance_2022,Fang_cryptocurrencytradingsurvey_2022,malik2022bitcoin}. The recent work \cite{takaishi_rough_2020} provided evidence that Bitcoin's volatility is rough, further justifying the inclusion of the rough volatility model in our work. The rapid increase in the number of listed coins on major cryptocurrency exchanges in recent years makes it more feasible to apply machine learning models such as the LSTM, and to demonstrate universality features. \\ 

\begin{figure}[H]
\centering
\includegraphics[width=0.9\textwidth]{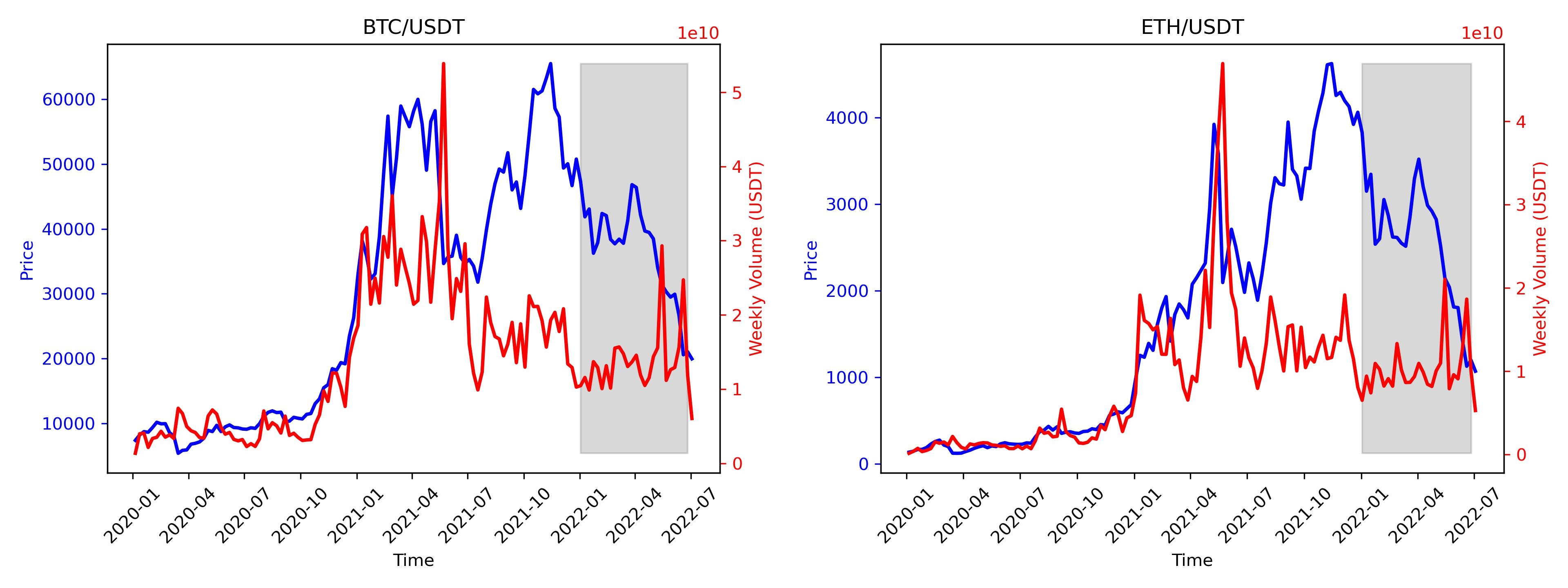}
\caption{Price and weekly USDT trading volume of BTC and ETH, with the period Jan to Jun 2022 shaded.}
\label{fig:btc_eth_chart}
\end{figure}

The year 2022 corresponds to a significant price and volume drop in cryptocurrencies compared to the peak in late 2021, or the beginning of a ``crypto-winter'' \footnote{Mccrank, J. (2022, February 14). Wall St Week Ahead Crypto investors face more uncertainty after rocky start to 2022. Reuters. https://www.reuters.com/business/finance/wall-st-week-ahead-crypto-investors-face-more-uncertainty-after-rocky-start-2022-2022-02-11/} \footnote{Harrison, E. (2022, May 10). The crypto winter is here. Bloomberg.com. https://www.bloomberg.com/news/newsletters/2022-05-10/the-crypto-winter-is-here} \footnote{Howcroft, E. (2022, June 13). Cryptocurrency market value slumps under \$1 trillion. Reuters. https://www.reuters.com/business/finance/cryptocurrency-market-value-slumps-under-1-trillion-2022-06-13/} \footnote{Coingecko. (2023). 2022 Annual Crypto Industry Report. CoinGecko. https://www.coingecko.com/research/publications/2022-annual-crypto-report}. For example, we display in Figure \ref{fig:btc_eth_chart} the prices and volumes from the two largest coins Bitcoin (BTC) and Ethereum (ETH).\\ 

There have been a few studies focusing on using machine learning to predict cryptocurrency volatility. In \cite{damato_deeplearningcryptovol_2022}, deep neural networks are used to model the volatilities of the three most popular cryptocurrencies. The papers \cite{kristjanpoller2018hybrid} and \cite{amirshahi2023hybrid} consider a hybrid approach combining neural networks and GARCH models to model cryptocurrency volatility, with the former using only Bitcoin and the latter using 27 coins. Compared with existing literature, we use a much larger set of 213 cryptocurrencies which enables us to show the universality of the volatility formation process across the asset class.\\

The rough volatility paradigm allows us to generate stylized facts of realized volatility \citep{gatheral_volatilityisrough_2018,rough1}. It is also effective in pricing and hedging derivatives \citep{bayer_pricing_2016,el_perfect_2018,el_roughening_2019,el_characteristic_2019}. With the simplest version of this approach, namely the rough fractional stochastic volatility (RFSV) model, volatility forecasts can be made with a simple formula that requires essentially one parameter, namely the Hurst parameter. The Hurst parameter is systematically around 0.1 for all different asset classes \citep{gatheral_volatilityisrough_2018,rough1,rough3}, giving a quasi-universal volatility model and suggesting universality in the volatility formation process across assets. In this paper, we also estimate the Hurst parameter for many cryptocurrencies and demonstrate the same observation.\\

In \cite{rosenbaum_universality_2022}, the authors presume that the universal volatility formation mechanism involves past returns. They extend the RFSV approach incorporating strong Zumbach effect \citep{zumbach_volatility_2010} which corresponds to the feedback effect of price trends on volatility. To do so they incorporate a quadratic rough Heston (QRH) component to the RFSV. We recall the QRH model is particularly popular as it is able to reproduce market observations of SPX and VIX options \citep{gatheral2020quadratic,rosenbaum_deep_2022} thanks to its rough nature and the Zumbach effect.\\


%
We focus on the problem of forecasting the daily volatility of cryptocurrencies one day ahead. In particular, one of our goals is to understand the universal aspects of the volatility formation process. One element towards this will be to investigate whether non-asset-dependent, generic, forecasting devices can lead to optimal predictions of future volatilities. First, we show that the LSTM approach performs better than classical autoregressive models, one of which is the heterogeneous autoregressive (HAR) model \citep{corsi_2009}. Then we use a combination of the RFSV and QRH for the parsimonious model candidate. We also conduct comparison tests to show that for this model, the universal version of the model performs very similarly to the one calibrated on each coin. This further confirms the universality of the volatility formation process from a parametric perspective. \\

We will follow the same structure as our companion paper \cite{rosenbaum_universality_2022}. In Section 2, we describe our data and model evaluation metrics. In Section 3, we describe both the parametric and non-parametric forecasting devices. We compare in Section 4 the performances of different forecasting models to provide evidence of a universal volatility formation mechanism for cryptocurrencies. Our approach also enables us to uncover asymmetry properties in the cryptocurrencies' volatilities in contrast to the stocks' volatilities. In Section 5, we aim to describe a universal mechanism for the volatility formation process with a parsimonious parametric model relating future volatility with past volatilities and returns. We conclude in Section 6.\\

\section{Data and evaluation metrics}

Our dataset contains 5-minute intraday prices of cryptocurrencies traded on the Binance exchange, with USDT, a popularly used stablecoin, as the base currency. Stablecoins are cryptocurrencies with price stabilization mechanisms, often in the form of pegging to some other assets such as traditional fiat money or other cryptocurrencies \citep{mita2019stablecoin}. In the case of USDT, it is pegged 1-to-1 with US dollars and can be considered a cryptocurrency version of USD\footnote{More about USDT can be found at https://tether.to/en/how-it-works.}. The Binance exchange is one of the largest cryptocurrency exchanges by trading volume and offers easily obtainable market data, including price, volume, and number of trades in intervals as granular as 5 minutes. The data is directly obtained from Binance.com.\\ 

We select 213 coins after filtering out coins that are stablecoins, leveraged, or have low liquidity or insufficient data. The training set is chosen to be the two-year period from \textit{2020-01-01} to \textit{2021-12-31}, and the test set is a six-month period from \textit{2022-01-01} to \textit{2022-06-30}.\\

\begin{figure}[H]
\centering
\includegraphics[width=0.5\textwidth]{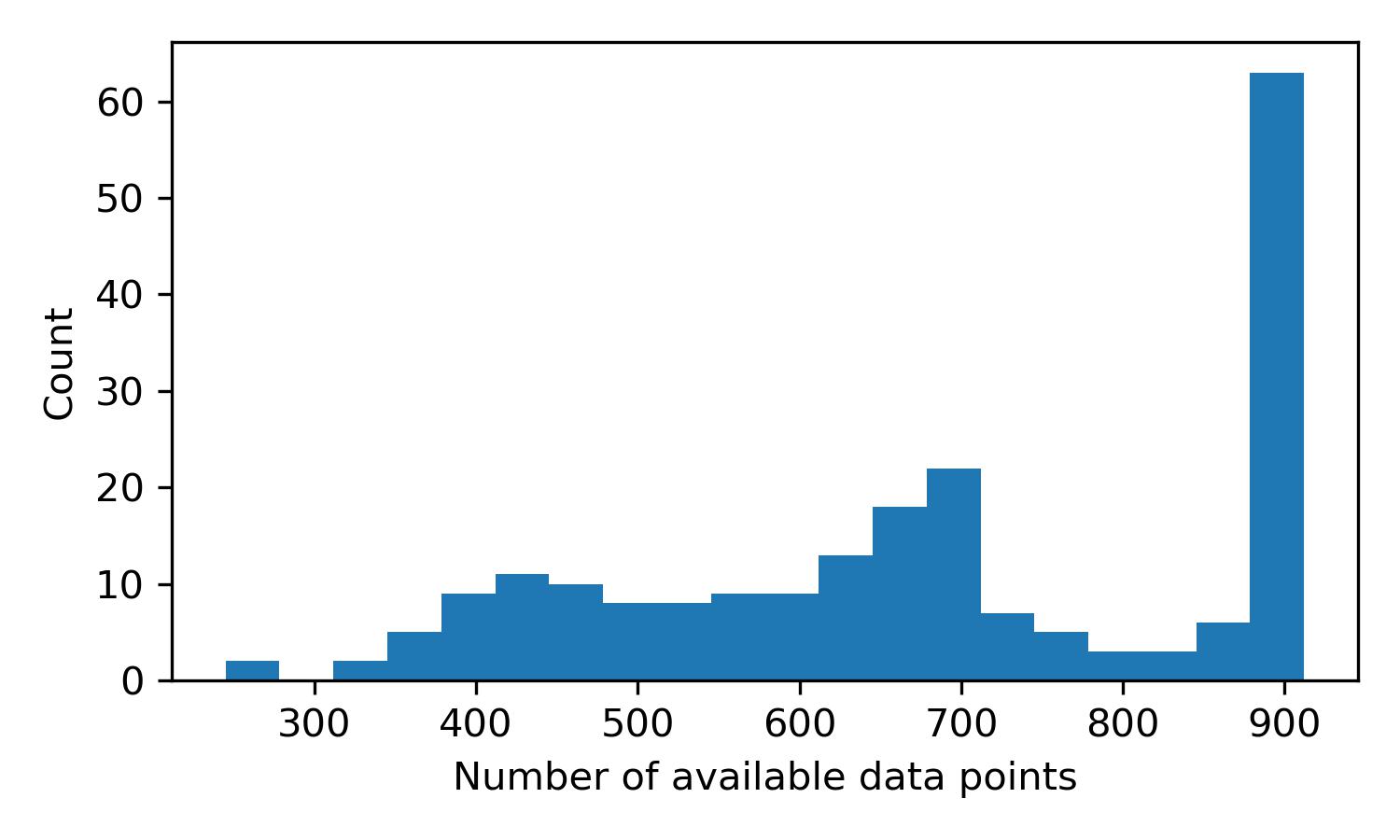}
\caption{Distribution of the number of available data points for selected coins.}
\label{fig:datapoint_dist}
\end{figure}

Due to the rapid increase in the number of cryptocurrencies in recent years and the issue of past data availability, we choose not to use training data from before 2020. The distribution of the number of available data points in the entire 2.5-year period for selected coins is given in Figure \ref{fig:datapoint_dist}. Note that the selected coins do not have the same number of available data points. While it would be ideal for every coin to have the same number of data points, it is not necessary for our approach and would require filtering out many more coins.\\

The daily realized volatility for day $t$, $\sigma_{t}$, the variable of interest, is computed as
\[\sigma_{t} = \sqrt{\sum_{i}{r_{t,i}^{2}}},\] where $r_{t,i}$  is the $i^{th}$ 5-minute returns of the day, defined as \[ r_{t,i} = \frac{P_{t,i}-P_{t,i-1}}{P_{t,i-1}},\] where $P_{t,i}$ is the $i^{th}$ 5-minute closing price of day $t$ ($P_{t,0}$ is the open price of day $t$). Similarly, the return of day $t$ is defined as \[ r_{t} = \frac{P_{t}-P_{t-1}}{P_{t-1}}, \] where ${P_{t}}$ is the closing price of day $t$. Since cryptocurrencies trade 24 hours per day every day, daily volatilities and returns are for 24-hour periods. The volatilities and returns are scaled and normalized respectively by the individual coin's statistics calculated on the training set.\\

For model evaluation, we choose to use the mean-squared-error or MSE, calculated as
\[MSE(\sigma,\hat{\sigma}) = \frac{1}{T}\sum^{T}_{t=1}(\hat{\sigma}_{t}-\sigma_{t})^{2},\] where $T$ is the number of trading days in the test set and $\hat{\sigma}_t$ is the model forecast for $\sigma_t$. According to \cite{patton2011volatility}, it is a robust and homogeneous metric for evaluating and comparing volatility models.

\section{Forecasting devices}
Here we introduce the parametric and non-parametric forecasting devices in this study.\\

\subsection{Parametric models}
First, we consider two auto-regressive type models. One is a classical autoregressive (AR) model, the other is a heterogeneous autoregressive (HAR) model \citep{corsi_2009}. Due to the fact that cryptocurrencies are traded every day, for the HAR, the windows are chosen to be 7 and 30 instead of 5 and 22 in the literature on traditional markets. More precisely we define:
\begin{itemize}
    \item AR\textsubscript{p}
        \[ \hat{\sigma}_{t} = \alpha_{0} + \sum_{j=1}^{p}{\beta_{j}\sigma_{t-j}}\]
        \textit{p} is chosen to be 7 or 30.
    \item HAR
        \[ \hat{\sigma}_{t} = \alpha_{0} + \beta_{1}\sigma_{t-1} + \beta_{2}\sum_{j=1}^{7}{\sigma_{t-j}}+\beta_{3}\sum_{j=1}^{30}{\sigma_{t-j}}\]
\end{itemize}
These two models are fitted on individual coin data.\\

For the RFSV, the log-volatility is modeled as a fractional Brownian motion ${W^{H}_{t}}$:

\[\textrm{d} \log \sigma_{t} = \nu \textrm{d} W^{H}_{t}, \nu>0 \]

The Hurst parameter $H$ and $\nu$ can be estimated using the method in \cite{gatheral_volatilityisrough_2018}.

As in \cite{gatheral_volatilityisrough_2018,rosenbaum_universality_2022}, the predictor for $H < \frac{1}{2}$ is given by:
\begin{align*}
\widehat{\log \sigma_{t}} &= \frac{\cos(H\pi)}{\pi}\int^{t-1}_{-\infty}\frac{\log\sigma_{s}}{(t-s+1)(t-s)^{H+\frac{1}{2}}}\mathrm{ds},\\
\hat{\sigma_{t}} &= c \exp(\widehat{\log \sigma_{t}}),\\
\end{align*}
where $c = \exp(\frac{\Gamma(\frac{3}{2}-H)}{2\Gamma(H+\frac{1}{2})\Gamma(2-2H)}\nu^{2})$. We will see in Section 5 that $c$ is empirically very close to 1, and the impact of $\nu^{2}$ on the forecast is marginal. $H$ is calibrated using the pooled data set with all the coins.

\subsection{Non-parametric forecasting with LSTM}
The recurrent neural network (RNN) is suitable for sequential learning tasks. Among types of RNNs, the Long Short-Term Memory (LSTM) neural network, introduced in \cite{hochreiter_lstm_1997}, is one of the most popular choices for predicting financial time series. \\

For each timestep $t$, the following computations are performed: 
\begin{align*}
i^{l}_{t}&=s(W^{l}_{ii}x^{l}_{t}+b^{l}_{ii}+W^{l}_{hi}h^{l}_{t-1}+b^{l}_{hi}),\\
f^{l}_{t}&=s(W^{l}_{if}x^{l}_{t}+b^{l}_{if}+W^{l}_{hf}h^{l}_{t-1}+b^{l}_{hf}),\\
g^{l}_{t}&=SiLU(W^{l}_{ig}x^{l}_{t}+b^{l}_{ig}+W^{l}_{hg}h^{l}_{t-1}+b^{l}_{hg}),\\
o^{l}_{t}&=s(W^{l}_{io}x^{l}_{t}+b^{l}_{io}+W^{l}_{ho}h^{l}_{t-1}+b^{l}_{ho}),\\
c^{l}_{t}&=f^{l}_{t}*c^{l}_{t-1}+i^{l}_{t}*g^{l}_{t},\\
h^{l}_{t}&=o^{l}_{t}*SiLU(c^{l}_{t})
\end{align*}
where the superscript indexes the layer, $c^{l}_{t}$, $h^{l}_{t}$ are the cell and hidden states, $x^{l}_{t}$ equals the $t$-th element of the input sequence ($\sigma_t$) or ($\sigma_t,r_t$) when $l = 1$ or $h^{l-1}_{t}$ for $l > 1$, $s(\cdot)$ is the sigmoid function, $*$ is the Hadamard product, $W^{l}_{.}$ and $b^{l}_{.}$ are parameters to learn from data. LSTMs use gates $i$, $f$, and $o$ to capture long-range dependence in the sequential data and to avoid the vanishing gradient problem of classical RNNs \citep{hochreiter_lstm_1997,goodfellow_deep_2016}.\\

We use LSTMs with input length \textit{p}, which are also chosen to be 7 or 30. The inputs can be volatilities or both volatilities and returns. We denote them by LSTM\textsubscript{p,var} and LSTM\textsubscript{p,ret} respectively. The LSTM consists of one LSTM layer and one dense output layer. The dimensions of $h^1_t$ for LSTM\textsubscript{p,var} and LSTM\textsubscript{p,ret} are chosen to be 2 and 4 respectively. The number of hidden LSTM layers is 1, followed by 1 dense output layer. These configurations are tuned using a cross-validation split within the training data. We use the SiLU, introduced in \cite{hendrycks_gelu_2020}, as the activation function.\\

Neural network fitting uses stochastic gradient descent and the results are noisy. Therefore we train 10 identically structured models, initialized with 10 different seeds. Lastly, we use the average of all their predictions as the final prediction. The optimizer we use is Adam, introduced in \cite{kingma_adam_2015}, with a learning rate of 1e-3. The LSTMs are trained using the pooled data set.\\

\section{Towards universality in the volatility formation process}
In this section, we compare different forecasting devices and show that we can achieve optimal performance with the universal ones. We also analyze the local sensitivities of the LSTMs and relate them to asymmetric volatility. \\

\subsection{Parametric vs LSTM}
\begin{figure}[H]
\centering
\includegraphics[width=0.8\textwidth]{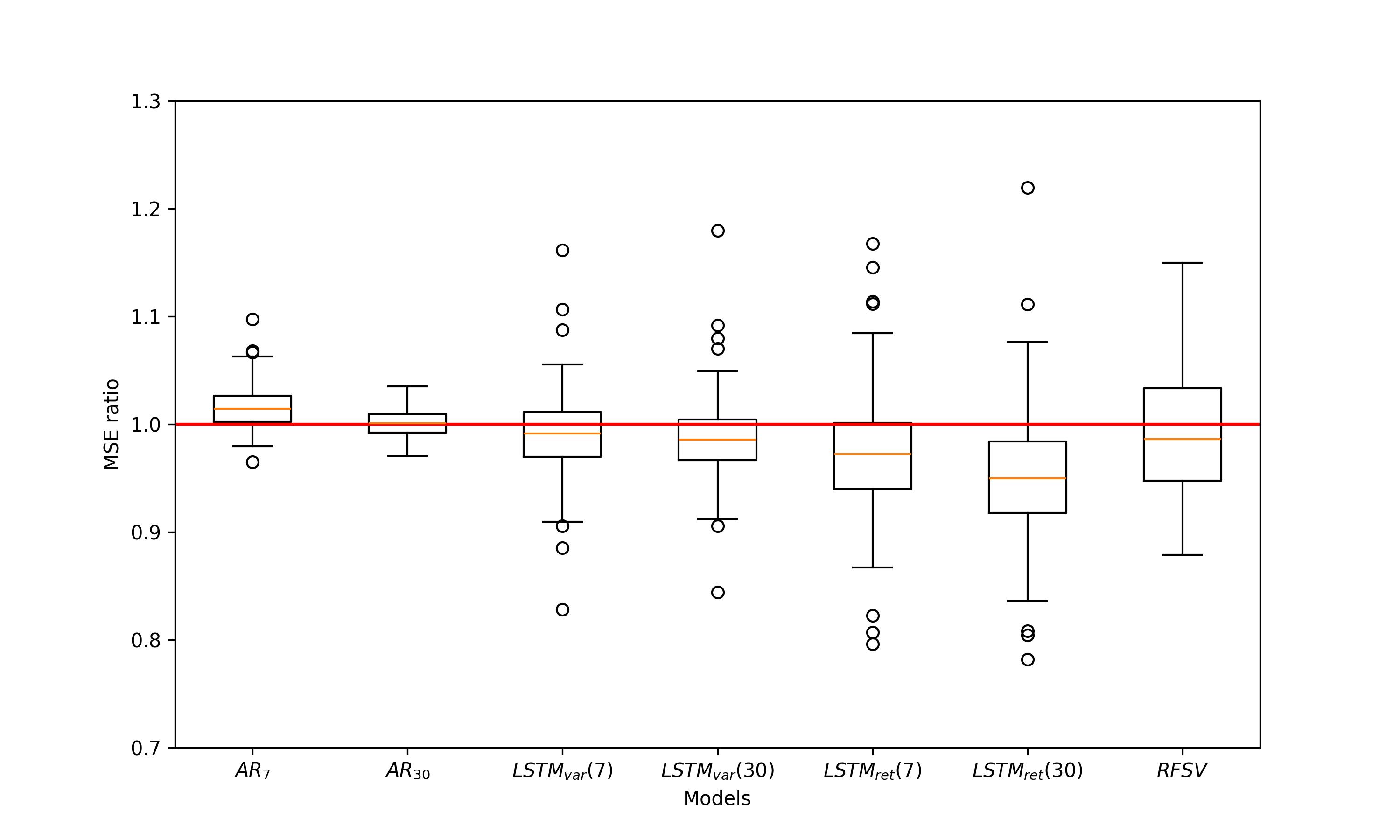}
\caption{Boxplots of MSE ratios of each model using HAR as the baseline.}
\label{fig:1}
\end{figure}

\noindent In Figure \ref{fig:1}, we have each model's MSE relative to HAR, grouped by coins. For each coin, we calculate the MSE ratio on a coin as the models' MSE divided by the baseline model's MSE on that coin's out-of-sample data. If the MSE ratio is less than 1 for a certain coin and model, then the model outperformed the baseline on the coin. We can see that:
\begin{itemize}
    \item AR\textsubscript{7} underperforms HAR.
    \item AR\textsubscript{30} is on par with HAR.
    \item LSTMs outperform HAR in general.
    \item LSTMs with a longer window show significantly better performances than those with a shorter window.
    \item LSTMs with return inputs show significantly better performances than those without.
    \item RFSV outperforms HAR, is slightly better than LSTM\textsubscript{7,var} and is close to LSTM\textsubscript{30,var}, but underperforms LSTM\textsubscript{7,ret} and LSTM\textsubscript{30,ret}.
    
\end{itemize}

These are largely consistent with the relative performances in \cite{rosenbaum_universality_2022}. We see that LSTMs and RFSV, which are calibrated with the pooled dataset, outperform linear models calibrated specifically to each coin. This not only supports the use of the two non-linear model types, but also the suspected universality in volatility which is captured by these models.\\

\subsection{Local sensitivities and asymmetric volatility}
\begin{figure}[]
\centering
\includegraphics[width=0.9\textwidth]{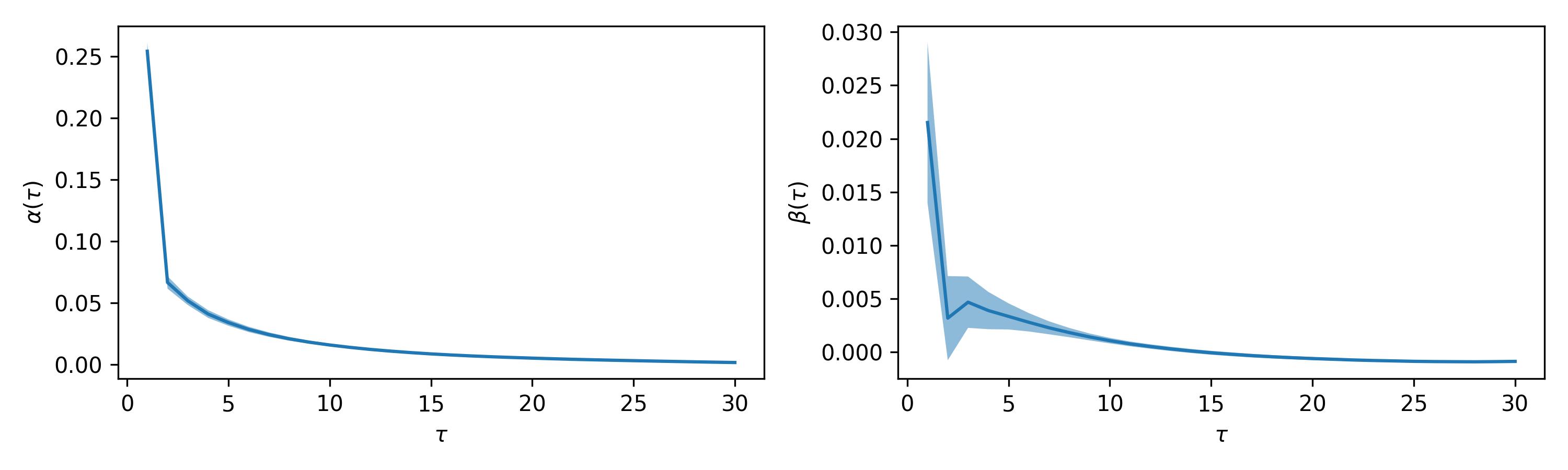}
\caption{Mean $\alpha(\tau)$ and $\beta(\tau)$ across coins and a one standard deviation band.}
\label{fig:2}
\end{figure}

\noindent We define local sensitivities by:
\[ \alpha_{t}(\tau) := \frac{\partial{\hat{\sigma}_{t}}}{\partial{\sigma^{2}_{t-\tau}}}, \:  \beta_{t}(\tau) := \frac{\partial{\hat{\sigma}_{t}}}{\partial{r_{t-\tau}}} \: \text{for} \: \tau = 1,2,...,30.\]
Here we obtain the gradients of the predictions with respect to each of the inputs of LSTM\textsubscript{30,ret}. For each coin, $\alpha(\tau)$ is given by the average of $\alpha_{t}(\tau)$ across $t$ in the test set, similarly $\beta(\tau)$ is the average of $\beta_{t}(\tau)$ across $t$ in the test set.
We then plot the average of these values across coins with a band of one standard deviation in Figure \ref{fig:2}. The plot of $\alpha(\tau)$ looks very similar to the one for stocks in \cite{rosenbaum_universality_2022}. We observe that the values are significantly positive, have a similar magnitude, and decay with increasing $\tau$.\\

For $\beta(\tau)$, the values are one order smaller than $\alpha(\tau)$, also decay in $\tau$, and are positive for cryptocurrencies. In \cite{rosenbaum_universality_2022}, $\beta(\tau)$ are observed to be negative for stocks. In both cases, the LSTMs captured a significant effect of returns on next-day volatilities, but the different signs of $\beta$ for the two asset classes are a curious observation. If we use volatility as an indicator of trading activity or even the popularity of an asset, we might hypothesize that cryptocurrency traders react to positive returns more than to negative returns, while stock traders do the opposite. By extension, we can also speculate that the typologies of market participants are different between the two asset classes.\\

It has been previously observed that for cryptocurrencies, positive price shocks increase the volatility more than negative shocks do \citep{baur_asymmetric_2018,cheikh_asymmetric_2020,kakinaka2022asymmetric}. This phenomenon is known as inverted asymmetric volatility. In traditional markets, such as the stock market, the observed asymmetry is reversed, \textit{i.e.} negative price shocks increase volatility more than positive shocks. Such an observation has often been attributed to the ``fear of missing out'' (FOMO) of uninformed cryptocurrency traders \citep{delfabbro2021psychology}, and the existence of ``pump and dump'' schemes \citep{li2021cryptocurrency}. A common conclusion in these previous works is that the observation comes from the difference in behaviour and trading activity by uninformed and informed traders dominating the cryptocurrency and stock market respectively.\\

 A nice byproduct of our methodology is a novel way to capture this inverted asymmetry in cryptocurrencies with the LSTM. We have also provided supportive evidence of this phenomenon across a much wider selection of cryptocurrencies than in previous studies of asymmetric volatility, confirming this observation on a large scale. Additionally, in tandem with \cite{rosenbaum_universality_2022}, in which an asymmetry is also observed using a similar method, we have strong evidence from large baskets of assets from both asset classes that the asymmetry in cryptocurrencies is opposite from that in the stock market.\\

\begin{figure}[]
\centering
\includegraphics[width = 0.6\textwidth]{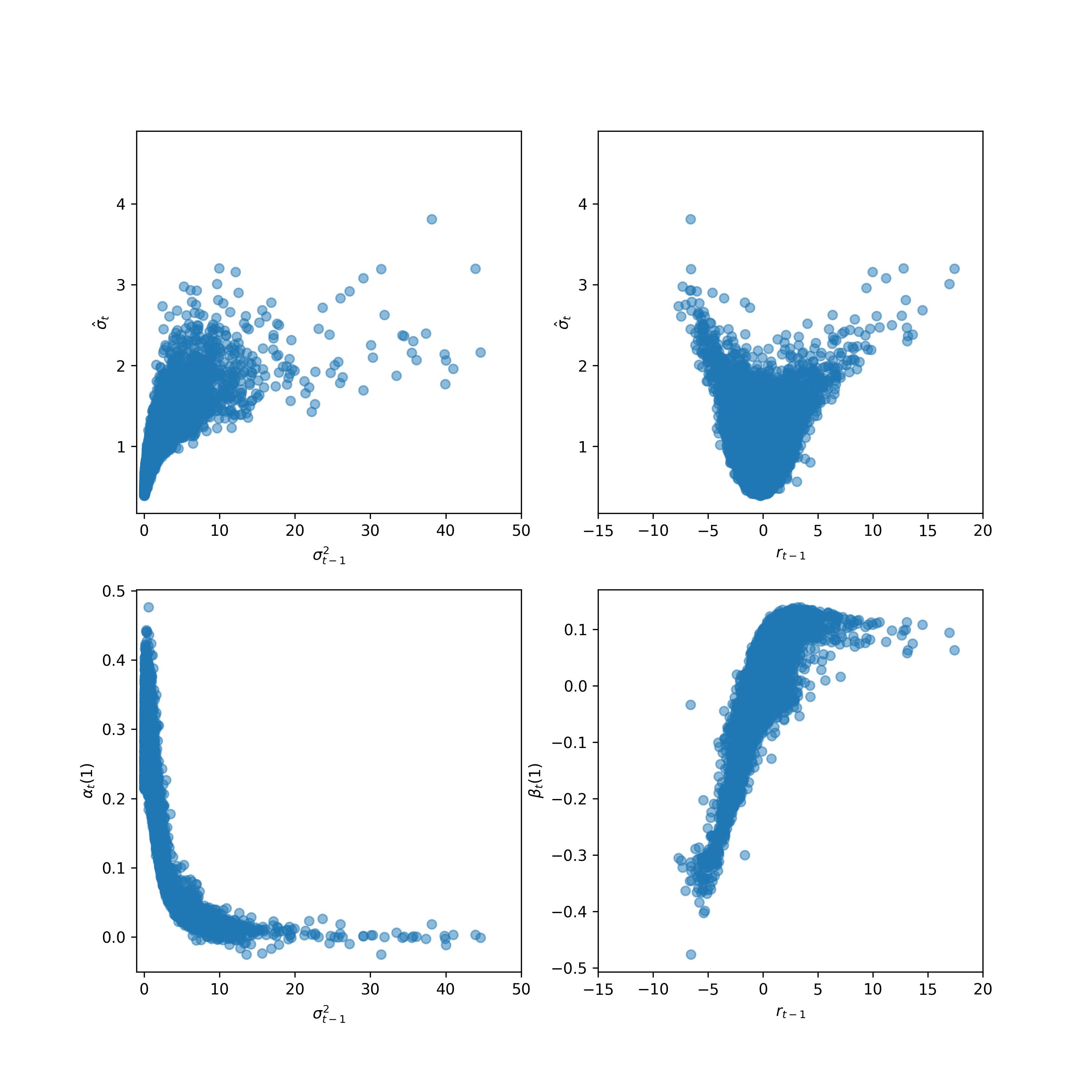}
\caption{Top left: $\hat{\sigma}$ against $\sigma^{2}_{t-1}$, top right: $\hat{\sigma}$ against $r_{t-1}$, bottom left: $\alpha_{t}(1)$ against $\sigma^{2}_{t-1}$, bottom right: $\beta_{t}(1)$ against $r_{t-1}$. }
\label{fig:alpha_beta}
\end{figure}

Figure \ref{fig:alpha_beta} plots the model prediction $\hat{\sigma}$ and gradients $\alpha_{t}(\tau)$, $\beta_{t}(\tau)$ against $\sigma_{t-1}^{2}$ and $r_{t-1}$ for all datapoints, without grouping by coin. 

\subsection{Comparing universal and non-universal LSTMs}
\subsubsection{Universality across different market capitalisation levels}
\begin{figure}[]
\centering
\includegraphics[width=0.6\textwidth]{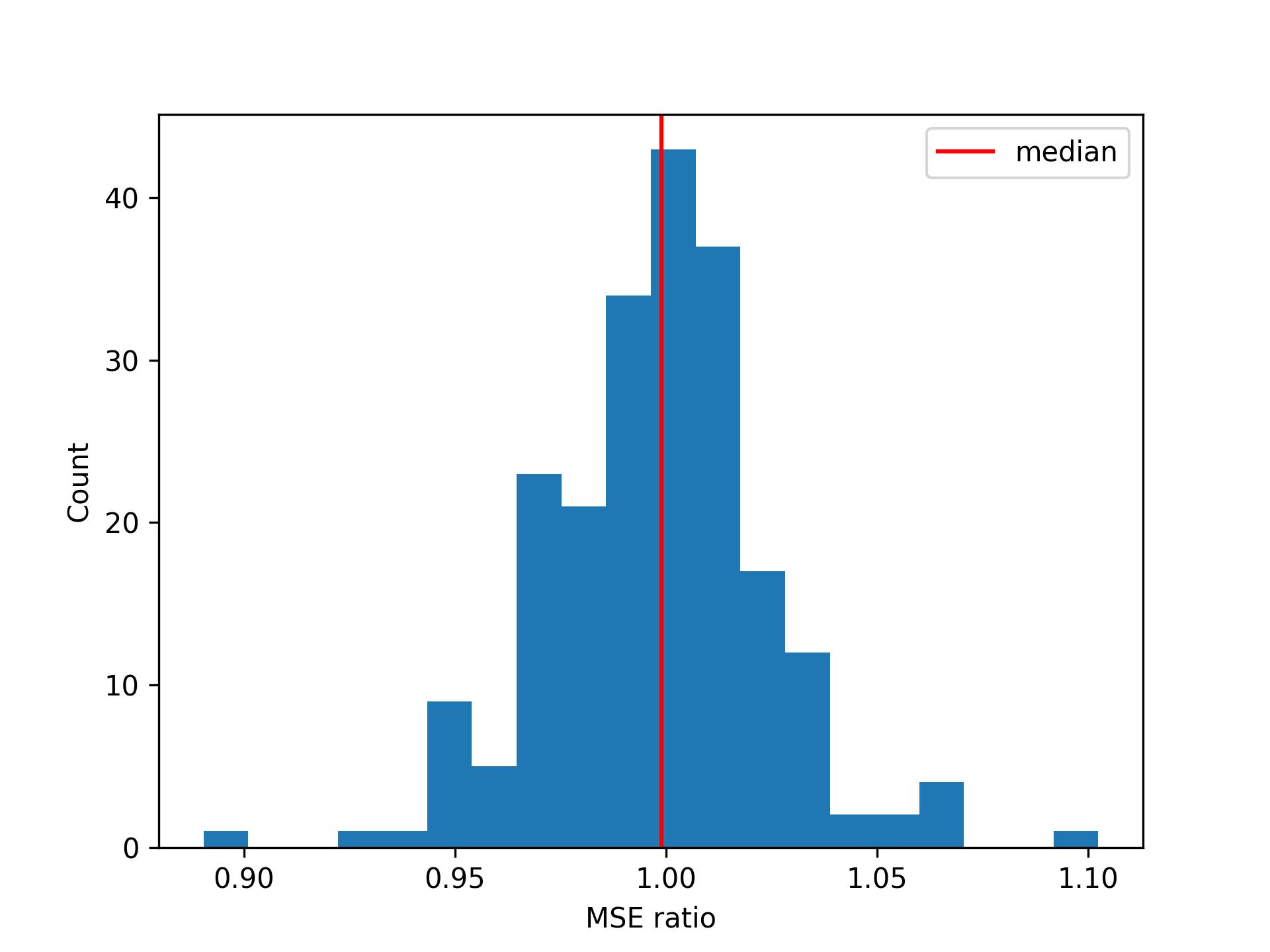}
\caption{Distribution of MSE ratios of LSTM\textsubscript{30,ret} trained with the top 50 coins by market capitalization using the universal LSTM\textsubscript{30,ret} as baseline.}
\label{fig:LSTM_ret_30_L50_vs}
\end{figure}

A new LSTM\textsubscript{30,ret} is trained on the top 50 coins by market capitalization (on 2021-12-31, the last day of the training period) only and compared with the universal one on the test period. The relative performance is shown in Figure \ref{fig:LSTM_ret_30_L50_vs}. There is no degradation of performance by the much smaller set of coins used in training, which suggests some universality of volatility formation across coins at different levels of market capitalization.\\

\subsubsection{No asset dependent mechanisms}
Next, the universal LSTM\textsubscript{30,ret} is fine-tuned to each individual coin's data to study whether this would improve its performance on individual coins.
Fine-tuning of neural networks is done by initiating a neural network's parameters with a trained one, then further training it with an often smaller dataset.
In our case, for each available coin, we initialize neural networks with the universal LSTM\textsubscript{30,ret}, then train it further with only that coin's training data. \\

\begin{figure}[H]
\centering
\includegraphics[width = 0.6\textwidth]{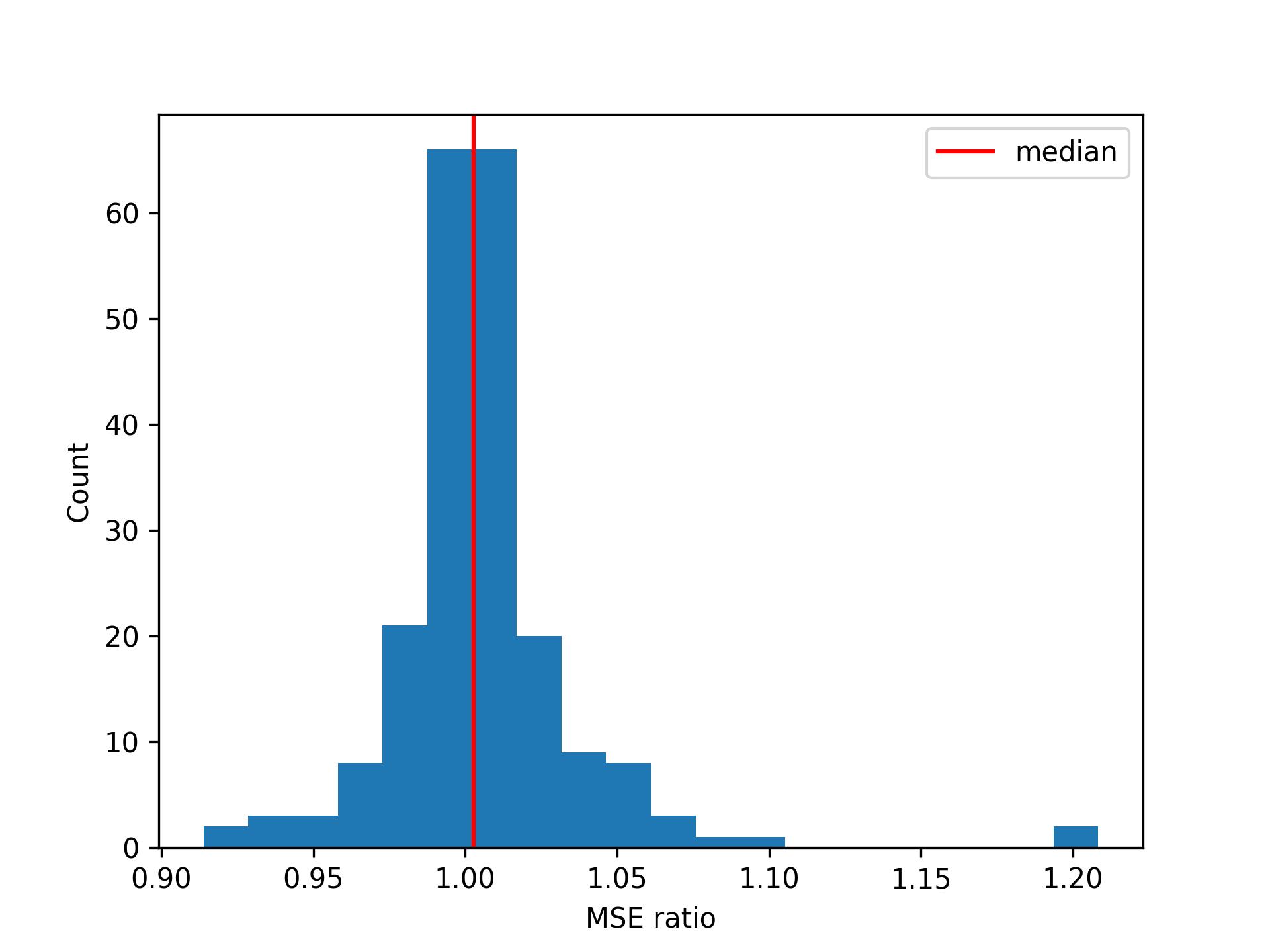}
\caption{Distribution of MSE ratios of LSTM\textsubscript{30,ret} fine-tuned to each coin using the universal LSTM\textsubscript{30,ret} as baseline.}
\label{fig:LSTM_ret_30_fine}
\end{figure}

If there were any asset-specific mechanisms that could be captured by the LSTM, we should see the fine-tuned LSTMs performing better. Here we allow all parameters to be changed during fine-tuning. The relative performance is shown in Figure \ref{fig:LSTM_ret_30_fine}. The fine-tuned models do not perform better. This supports that there are no asset-dependent mechanisms in volatility formation for the coins.

\section{Uncovering universal mechanisms in the volatility formation process}

In this section, we calibrate the RFSV to each coin to confirm the universality of the volatility formation process. Then we incorporate the strong Zumbach effect into the rough-volatility based model by adding a quadratic rough Heston component. By comparing the combined model to the LSTM and RFSV, we obtain a promising candidate for the description of a universal volatility formation process.\\

\subsection{RFSV}
\begin{figure}[H]
\centering
\includegraphics[width=0.6\textwidth]{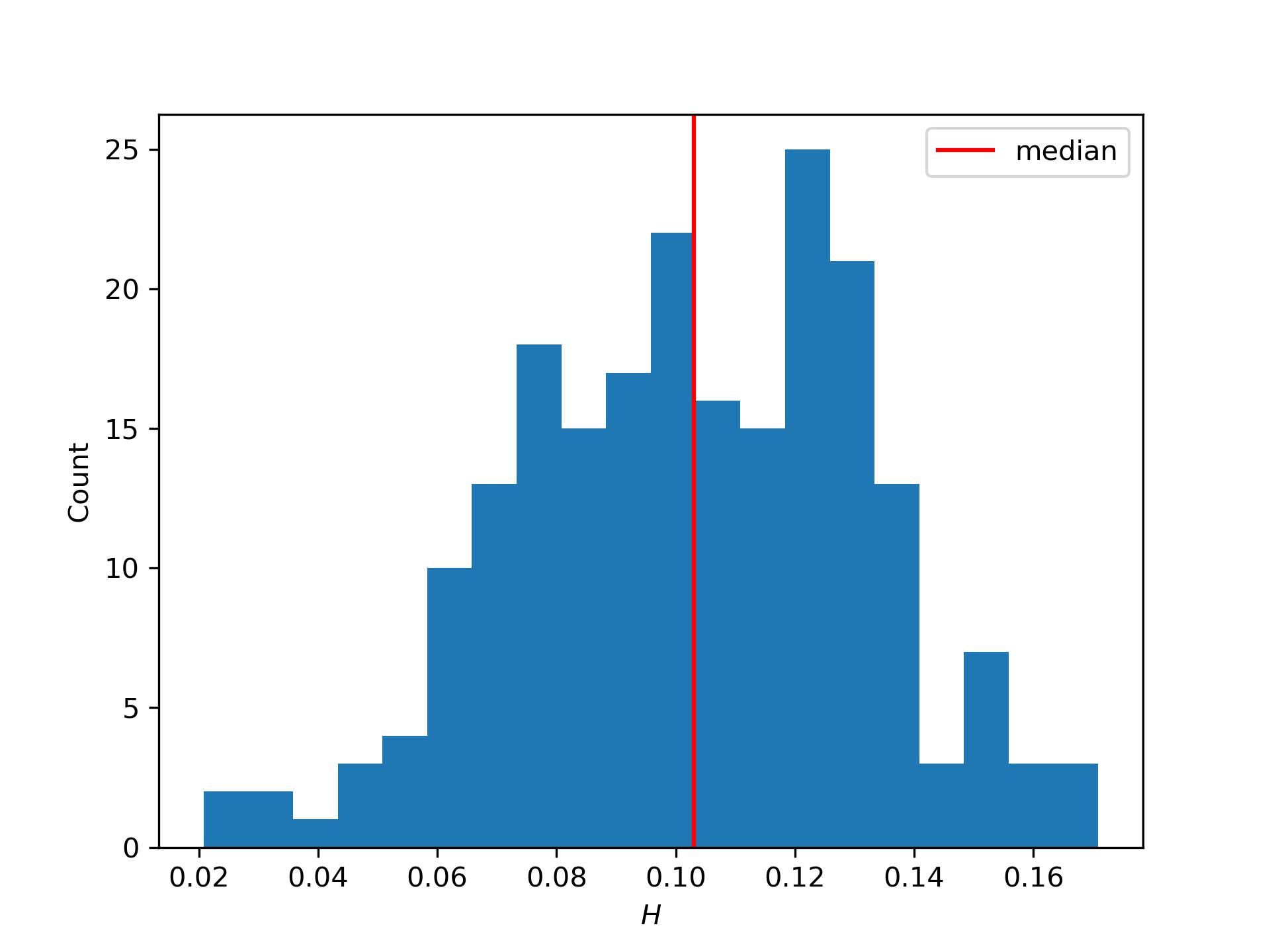}
\caption{Distribution of H.}
\label{fig:Hdist}
\end{figure}

\begin{figure}[H]
\centering
\includegraphics[width=0.6\textwidth]{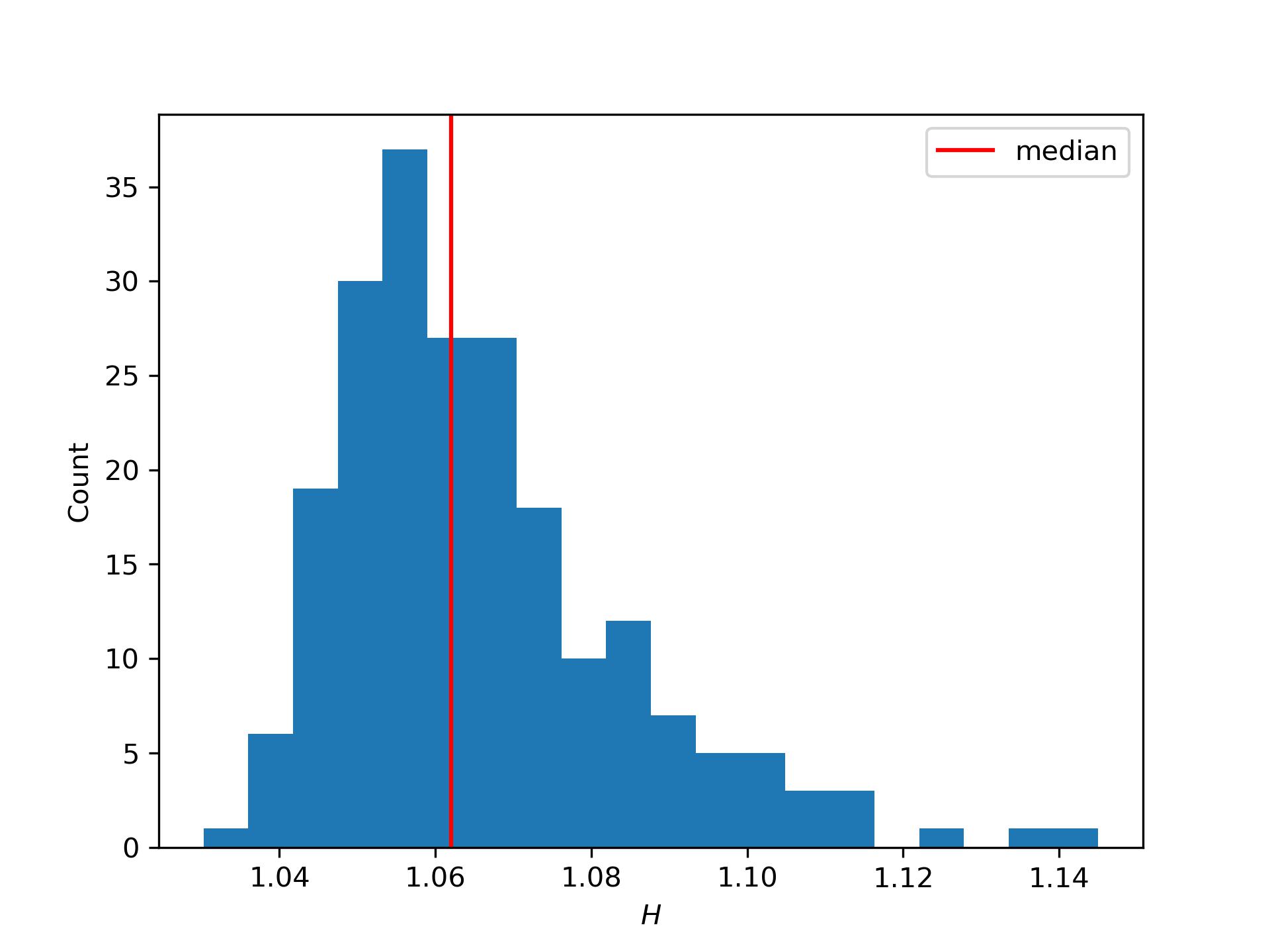}
\caption{Distribution of c.}
\label{fig:cdist}
\end{figure}

For RFSV, we first estimate H and c for each coin, and the distributions are shown in Figures \ref{fig:Hdist} and \ref{fig:cdist}. H and c tend to be higher than the ones for stocks as found in \cite{rosenbaum_universality_2022}. The median of H and c are 0.103 and 1.06 respectively.\\

\begin{figure}[H]
\centering
\includegraphics[width = 0.6\textwidth]{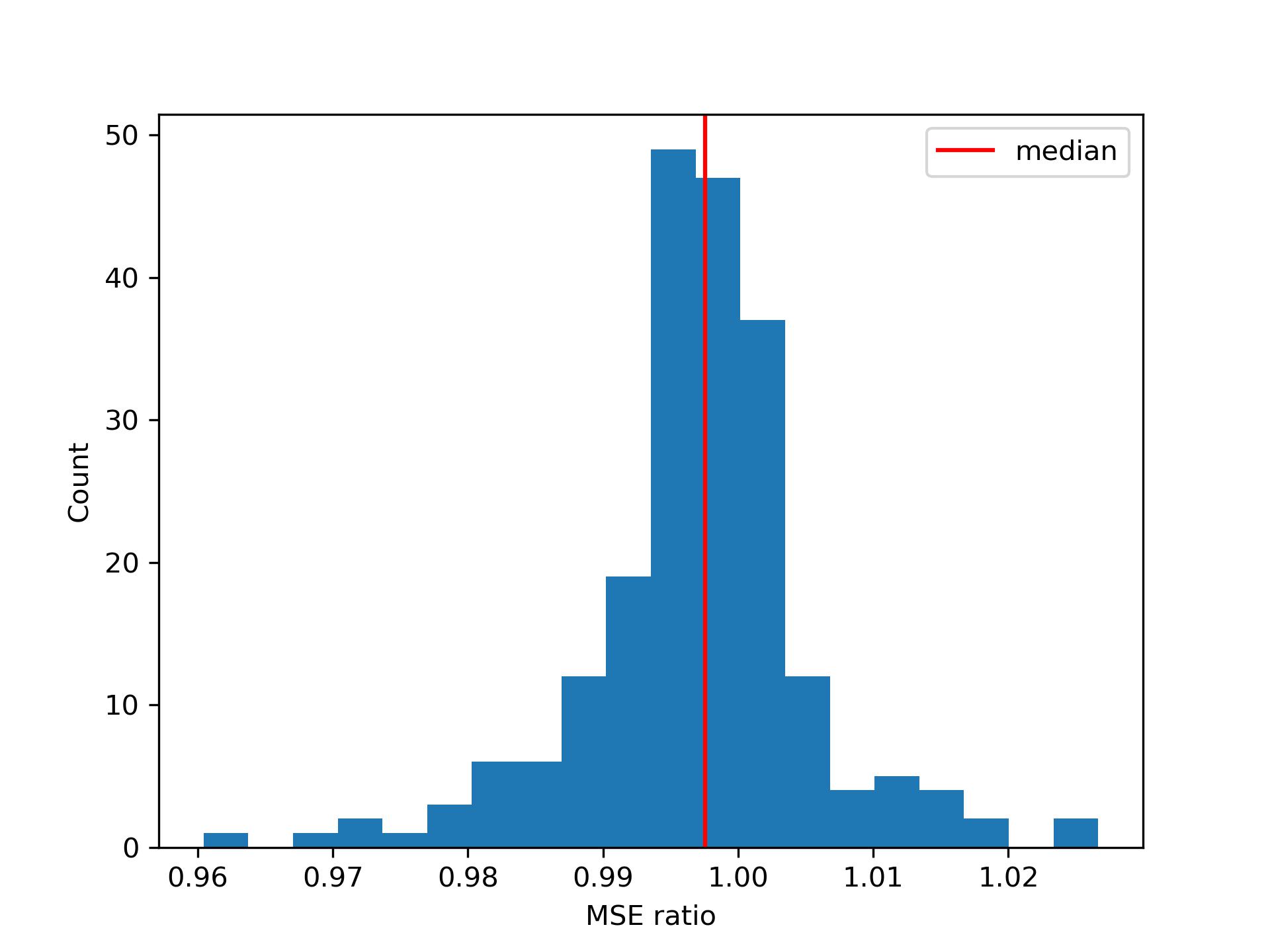}
\caption{Distribution of MSE ratios of RFSV calibrated to each coin using the universal RFSV as baseline.}
\label{fig:7}
\end{figure}

To once again check for universality from a parametric perspective, RFSVs are calibrated to each coin and compared with the universal RFSV that is calibrated to all coins. The relative performance is shown in Figure \ref{fig:7}. The asset-specific RFSVs perform very similarly to the universal RFSV. Similar to fine-tuned LSTMs, we would expect improvements from the asset-specific RFSVs if there were any coin-specific mechanisms that could be captured by RFSV. Once again we obtain further evidence of a lack of asset-specific mechanisms from a parametric perspective.

\subsection{RFSV enhanced with QRH}

Here we use the forecasting device based on the Zumbach effect from \cite{rosenbaum_deep_2022,rosenbaum_universality_2022} via a quadratic rough Heston component. The forecasting formula from the QRH model is given by 
\begin{align*}
\hat{\sigma}^2_t &= a(Z_{t-1} - b)^2 + c,\\
Z_t &= \sum_{i=1}^{n} c^d_i Z_{i,t},\\
Z_{i,t} &= e^{- \gamma ^d_i} Z_{i,t-1} + r_t, \quad Z_{i,0} = z_{i,0}, \quad i = 1, \ldots, n,
\end{align*}
where $a, b, c > 0$, and $(c^d_i, \gamma ^d_i)_{i=1,\ldots,n}$ are given by the same multi-factor approximation of the rough kernel function $K(t):= \frac{t^{H-1/2} }{\Gamma(H + 1/2)}$ \citep{rosenbaum_deep_2022}. Given $H$ and $n$, $(c^d_i, \gamma^d_i)_{i=1,2,\dots,n}$ are not free parameters. The predictions are essentially moving averages of past realized volatilities and are not sensitive to the earliest inputs due to the exponential decay in weights. In practice after computing $Z_{i,t}$, one needs to discard the earliest samples without sufficient histories. By including the QRH, we have three more parameters $a,b,c$ to calibrate which can be done by regressing $\sigma^2_t$ on $(Z_{t-1},Z^2_{t-1})$.\\

\begin{figure}[]
\centering
\includegraphics[width = 0.8\textwidth]{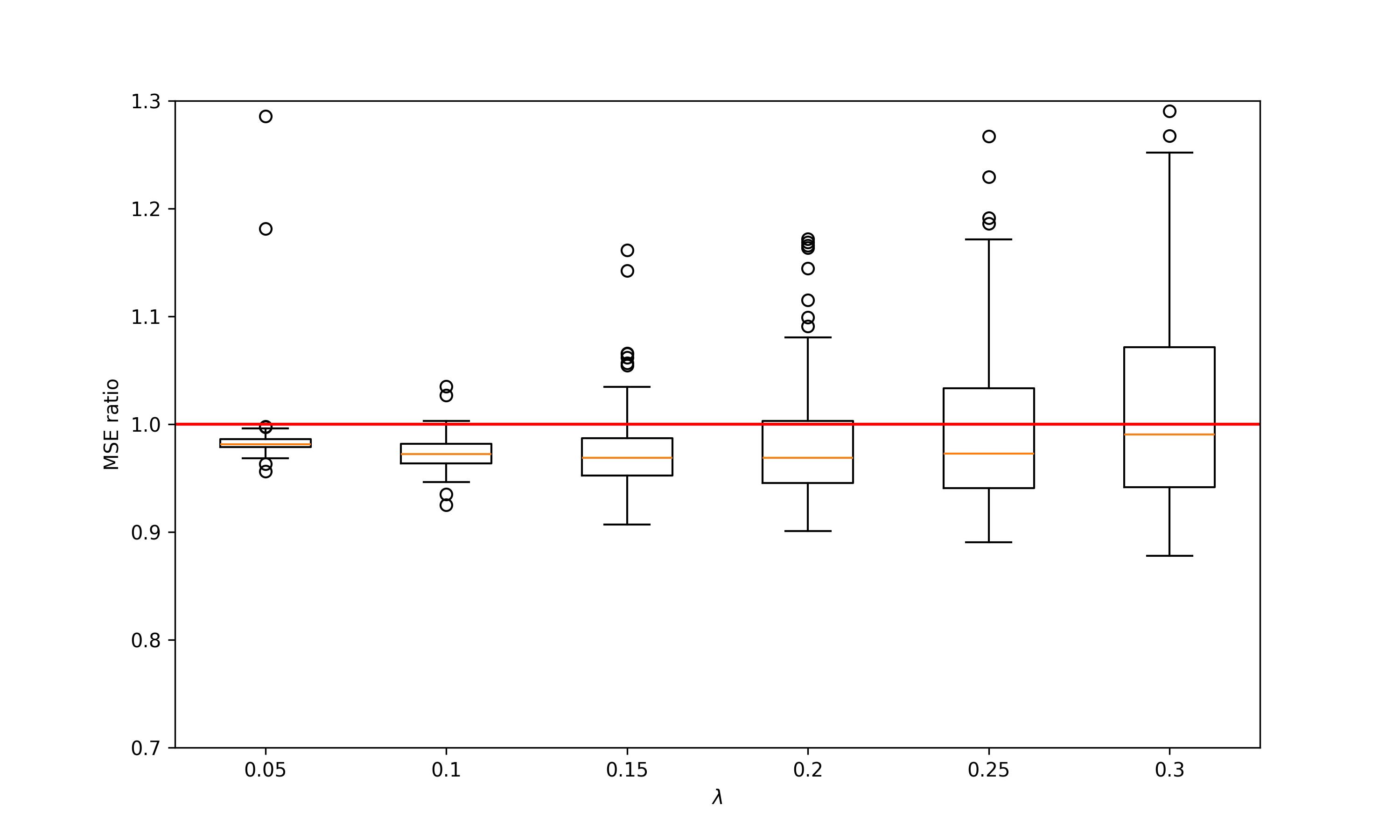}
\caption{Boxplots of MSE ratios of $(1-\lambda)\hat{\sigma}^{RFSV} + \lambda\hat{\sigma}^{QRH}$ with different positive values for $\lambda$ using RFSV only ($\lambda = 0)$ as baseline.}
\label{fig:RFSV_QRH}
\end{figure}

\begin{figure}[]
\centering
\includegraphics[width = 0.8\textwidth]{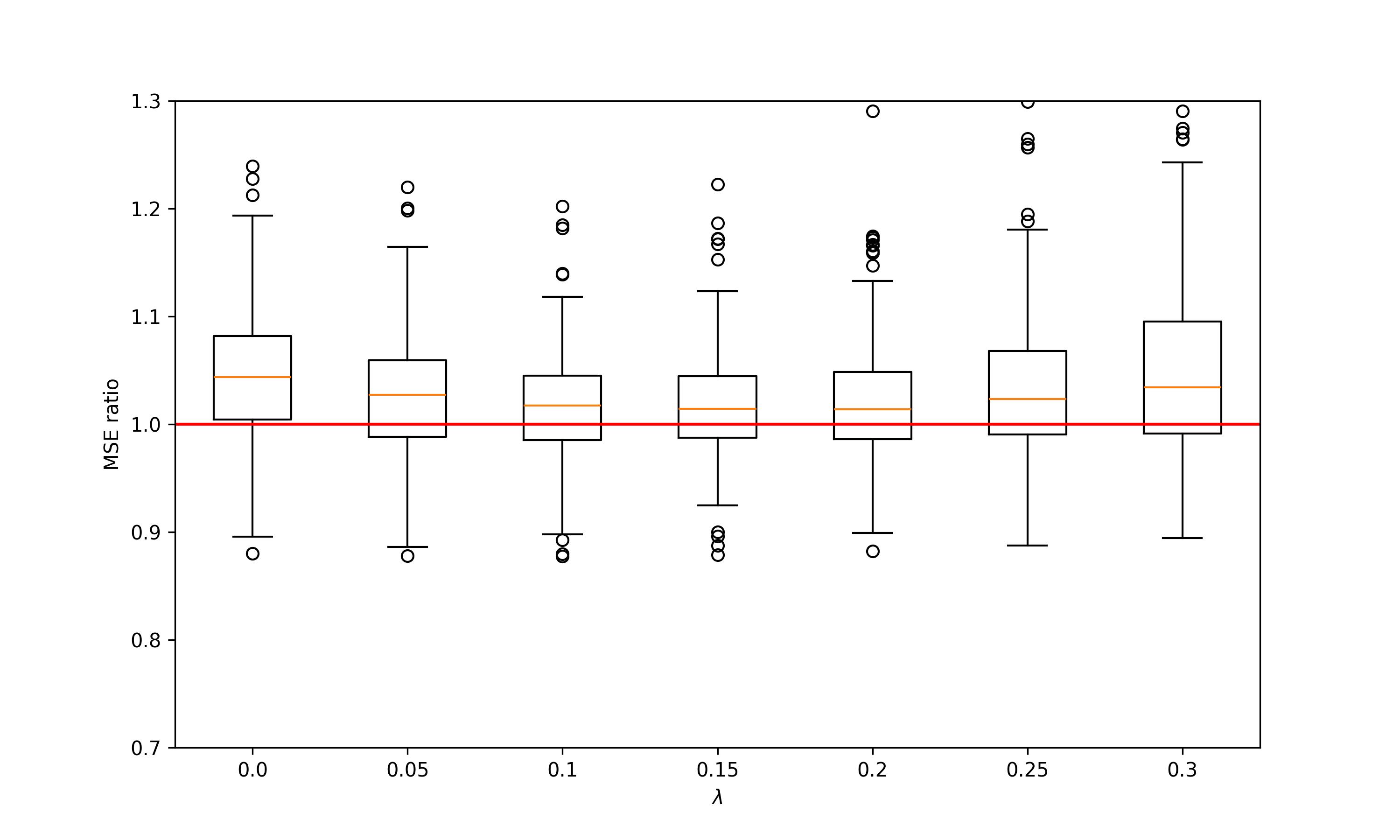}
\caption{Boxplots of MSE ratios of $(1-\lambda)\hat{\sigma}^{RFSV} + \lambda\hat{\sigma}^{QRH}$ with different values for $\lambda$ using LSTM\textsubscript{30,ret} as baseline.}
\label{fig:RFSV_QRH_vs_LSTM}
\end{figure}

We will combine the predictions from RFSV and QRH linearly. The final prediction is given by $(1-\lambda)\hat{\sigma}^{RFSV} + \lambda\hat{\sigma}^{QRH}$. The relative performances for different values for $\lambda$ to RFSV only are shown in Figure \ref{fig:RFSV_QRH}, the relative performances to LSTM\textsubscript{30,ret} are shown in Figure \ref{fig:RFSV_QRH_vs_LSTM}. \\

Incorporating past returns in the RFSV through the QRH component is a clear improvement. At $\lambda = 0.15$, the model performs best and is very similar to LSTM\textsubscript{30,ret}. This gives us a parsimonious parametric model candidate for a universal volatility mechanism.\\

We also note the larger optimal value for $\lambda $ than that in \cite{rosenbaum_universality_2022} for stocks. Past returns play a larger part in the volatility formation in cryptocurrencies than in stocks. \\ 

\section{Conclusion}
In this work, we adapt the volatility forecasting framework of \cite{rosenbaum_universality_2022} and extend it to a different asset class. The framework and forecasting devices used in the original paper are applied to the daily volatilities of cryptocurrencies at the start of the ``crypto-winter'' crisis where prices and trading volume dropped. This is also the first study of cryptocurrency volatility in which a wide selection of coins is used. In general, we find the framework to work very well in producing relevant universal volatility forecasting devices, as well as evidence of universal mechanisms in the volatility formation process, as was found in stocks. The findings of the applied methods, regarding predicting volatility, are thus remarkably largely similar in the very different datasets and contexts.\\

Using classical auto-regressive models, LSTM and RFSV, we show that the volatility during this period can be modeled and predicted well with the latter two. The non-parametric LSTM and parsimonious RFSV, like for US stocks, outperform all other traditional models used in industry and past literature. While constructing the RFSV forecaster, we estimate the Hurst parameters of the log volatilities for all the coins and showed that their volatilities are indeed ``rough''. 
While comparing the fine-tuned LSTM and RFSV to their universal counterparts, we observe a universality of this process across the wide selection of coins, which is also observed in stocks.\\

Next, we use the combination of RFSV and QRH to obtain a parsimonious model. We then find that the RFSV+QRH model can perform as well as the LSTM with both volatility and returns as input. This suggests that the main features in the volatility formation for cryptocurrencies can be described by rough volatility boosted with a strong Zumbach effect. This model has also once again confirmed the universality of the volatility formation process across coins. The obtained models and volatility characteristics can further our understanding of this young and evolving market. Remarkably, this device is found to also perform very well for stock volatility, suggesting a higher level of universality across asset classes.\\

In addition, the use of LSTM offers a new way to capture the effect known as asymmetric volatility. By analyzing the gradients of the volatility forecast with respect to past returns, we can capture and confirm the asymmetrical effect of past returns on volatility, and compare this effect with other asset classes. This effect can be further used to infer the behavior and informedness of the traders in these markets. Based on existing results, we can hypothesize cryptocurrency traders tend to react to positive shocks more and might be less informed compared to traditional market participants.\\


\section*{Acknowledgements}
\addcontentsline{toc}{section}{Acknowledgements}

Siu Hin Tang is supported by the SINGA award by A*Star Singapore. Mathieu Rosenbaum is supported by the École Polytechnique's chairs Deep finance and statistics and Machine learning and systematic methods. Chao Zhou is supported by the Ministry of Education in Singapore under the MOE AcRF grants A-0004255-00-00, A-0004273-00-00, A-0004589-00-00 and by Iotex Foundation Ltd under the grant A-8001180-00-00. 

\bibliographystyle{apalike}
\bibliography{biblio}

\end{document}